\documentclass[prb,preprint]{revtex4}
\usepackage{graphicx}
\begin{document}
\draft
\begin{abstract}
Starting from a model of an elastic medium,  partial differential
equations with the form of the coupled Einstein-Dirac-Maxwell
equations are derived. The form of these equations describes
particles with mass and spin coupled to electromagnetic and
gravitational type of interactions. A two dimensional version of
these equations is obtained by starting with a model in three
dimensions and deriving equations for the dynamics of the lowest
fourier modes assuming one dimension to be periodic. Generalizations
to higher dimensions are discussed.
\end{abstract}
\pacs{PACS numbers: } \vspace{.5in}
\title{A Derivation of the Classical Einstein-Dirac-Maxwell Equations From a Model of an Elastic Medium}
\author{John Baker}
\affiliation{ 2221 Parnassus Ct; Hayward, CA 94542} \maketitle

\section{Introduction} Dirac's equation describes the behavior of
particles with mass and spin and how they couple to the
electromagnetic field.  The usual form of Dirac's equation (with
$\hbar$, and $c$ set to unity) is\cite{ref:Sakurai}

\[ (\gamma^{\mu}\partial_{\mu}+m)\Psi(x)=0
\]

where $\Psi$ is a spinor field, $m$ is the particle mass and the
gamma matrices $\gamma^\mu$ satisfy
$\{\gamma^{\mu},\gamma^{\nu}\}=2I \delta_{\mu\nu}$.

 The electromagnetic field is
introduced by the minimal coupling prescription\cite{ref:Peskin}
$\partial_{\mu}\rightarrow D_{\mu}$, with

\[D_{\mu}=\partial_{\mu}+\imath e A_{\mu}(x)
\]

where $A_{\mu}$ is the electromagnetic vector potential.  Dirac's
equation can be further coupled to gravity (at the classical level)
and the equation then takes the
form\cite{ref:Finster,ref:Brill_Wheeler,ref:Brill_Cohen,ref:Smoller_Finster,ref:Smoller_Finster2}
\begin{equation}
\label{eq:full_dirac}
\tilde{\gamma}^{\mu}[\partial_{\mu}-\Gamma_{\mu}+\imath e
A_\mu]\Psi(x)+m\Psi(x)=0
\end{equation}
where $\Gamma_\mu$ is known as the spin connection, $A_\mu$ is the
electromagnetic vector potential, and $m$ is the mass. The
gravitational coupling enters through the modified dirac matrices
$\tilde{\gamma}^{\mu}$ which satisfy the anticommutation relation

\begin{equation}
\{\tilde{\gamma}^{\mu},\tilde{\gamma}^{\nu}\}=2I g^{\mu\nu}.
\end{equation}
  and the spin connection satisfies the additional constraint\cite{ref:Brill_Wheeler,ref:Brill_Cohen}
\begin{equation}
\label{eq:auxiliary} \frac{\partial \gamma'^\mu}{\partial
x^\nu}+\gamma'^\beta\Gamma^\mu_{\beta\nu}-\Gamma_\nu\gamma'^\mu+\gamma'^\mu\Gamma_\nu=0
\end{equation}
 where $\Gamma^\mu_{\beta\nu}$ are the usual Christoffel symbols.

The above form of Dirac's equation describes the dynamics of the
spinor field $\Psi$ when coupled to the scalar fields $A_{\mu}$ and
gravity.  There are two additional equations which describe the
dynamics of $A_{\mu}$ and $g_{\mu\nu}$, these are the Einstein field
equation

\begin{equation}
\label{eq:einstein_field}
 R_{\mu\nu}-\frac{1}{2}R=T_{\mu\nu}
\end{equation}

and Maxwell's equations

\begin{equation}
\label{eq:maxwell}
 \nabla_\mu F^{\mu\nu}=4\pi
e\bar{\Psi}\gamma^\nu\Psi
\end{equation}

The Equations~(\ref{eq:full_dirac}), (\ref{eq:einstein_field}) and
(\ref{eq:maxwell}) are collectively known as the
Einstein-Dirac-Maxwell
equations\cite{ref:Smoller_Finster,ref:Smoller_Finster2,ref:Krori}.
This paper will show that there exists a "Dirac-like" equation, with
the form of Equation~(\ref{eq:full_dirac}), that determines the
dynamics of the lowest Fourier modes of an elastic medium  and
furthermore the dimensional reduction involved in the Fourier
transform implies Equations~(\ref{eq:einstein_field}) and
(\ref{eq:maxwell}).

\section{Elasticity Theory}
\label{sec:elasticity_theory}
 The theory of elasticity is usually
concerned with the infinitesimal deformations of an elastic
body\cite{ref:Love,ref:Sokolnikoff,ref:Landau_Lifshitz,ref:Green_Zerna,ref:Novozhilov}.
We assume that the material points of a body are continuous and can
be assigned a unique label $\vec{a}$. For definiteness the elastic
body can be taken to be a three dimensional object so each point of
the body may be labeled with three coordinate numbers $a^{i}$ with
$i=1,2,3$.

If this three dimensional elastic body is placed in a large ambient
three dimensional space then the material coordinates $a^{i}$ can be
described by their positions in the 3-D fixed space coordinates
$x^{i}$ with $i=1,2,3$.  In this description the material points
$a^{i}(x^1,x^2,x^3)$ are functions of $\vec{x}$. A deformation of
the elastic body results in infinitesimal displacements of these
material points. If before deformation, a material point $a^0$ is
located at fixed space coordinates $x^{01},x^{02},x^{03}$ then after
deformation it will be located at some other coordinate
$x^1,x^2,x^3$.  The deformation of the medium is characterized at
each point by the displacement vector
\[u^i=x^i-x^{0i}
\]
which measures the displacement of each point in the body after
deformation.

It is our aim to take this model of an elastic medium and derive
from it equations of motion that have the same form as Dirac's
equation. In doing so we have to distinguish between the intrinsic
coordinates of the medium which we will call "internal" coordinates
and the fixed space coordinates which facilitates our derivation of
the equations of motion. In the undeformed state we may take the
external coordinates to coincide with the material coordinates $a^i=
x^{0i}$. The approach that we will use in this paper is to derive
equations of motion using the fixed space coordinates and then
translate this to the internal coordinates of our space.

We first consider the effect of a deformation on the measurement of
distance.  After the elastic body is deformed, the distances between
its points changes as measured with the fixed space coordinates. If
two points which are very close together are separated by a radius
vector $dx^{0i}$ before deformation, these same two points are
separated by a vector $dx^i=dx^{0i}+du^i$ afterwards. The square
distance between the points before deformation is then
$ds^2=(dx^{01})^2+(dx^{02})^2+(dx^{03})^2$. Since these coincide
with the material points in the undeformed state, this can be
written $ds^2=(da^1)^2+(da^2)^2+(da^3)^2$. The squared distance
after deformation can be written\cite{ref:Landau_Lifshitz}
$ds'^{2}=(dx^1)^2+(dx^2)^2+(dx^3)^2=\sum_i
(dx^i)^2=\sum_i(da^i+du^i)^2$. The differential element $du^i$ can
be written as $du^i=\sum_i\frac{\partial u^i}{\partial a^k }da^k$,
which gives for the distance between the points

\begin{eqnarray*}
ds'^2&=&\sum_i\left(da^i + \sum_k\frac{\partial u^i}{\partial
a^k}da^k\right) \left(da^i + \sum_l\frac{\partial u^i}{\partial
a^l}da^l\right)\\
&=&\sum_i\left(da^i da^i + \sum_k\frac{\partial u^i}{\partial
a^k}da^k da^i+ \sum_l\frac{\partial u^i}{\partial a^l}da^i da^l +
\sum_k\sum_l\frac{\partial u^i}{\partial a^k}\frac{\partial
u^i}{\partial a^l}da^k da^l\right)\\
&=&\sum_i\sum_k\left(\delta_{ik}+\left(\frac{\partial u^i}{\partial
a^k}+\frac{\partial u^k}{\partial a^i}\right)+\sum_l\frac{\partial
u^l}{\partial
a^i}\frac{\partial u^l}{\partial a^k}\right) da^i da^k\\
&=&\sum_{ik}\left(\delta_{ik}+2\epsilon'_{ik}\right)da^i da^k
\end{eqnarray*}
 where $\epsilon'_{ik}$ is
\begin{equation}
\label{eq:strain_tensor}
\epsilon'_{ik}=\frac{1}{2}\left(\frac{\partial u^i}{\partial
a^k}+\frac{\partial u^k}{\partial a^i}+\sum_l \frac{\partial
u^l}{\partial a^i}\frac{\partial u^l}{\partial a^k}\right)
\end{equation}
 The
quantity $\epsilon'_{ik}$ is known as the strain tensor. It is
fundamental in the theory of elasticity. In most treatments of
elasticity it is assumed that the displacements $u^i$ as well as
their derivatives are infinitesimal so the last term in
Equation~(\ref{eq:strain_tensor}) is dropped.  This is an
approximation that we will not make in this work.

The quantity
\begin{eqnarray}
\label{eq:metric_tensor}
 g_{ik}&=&\delta_{i,k}+\frac{\partial u^i}{\partial
a^k}+\frac{\partial u^k}{\partial a^i}+\sum_l \frac{\partial
u^l}{\partial a^i}\frac{\partial u^l}{\partial a^k}\\
&=&\delta_{i,k}+2\epsilon'_{ik}\nonumber
\end{eqnarray}
 is the metric for our system and
determines the distance between any two points.

That this metric is simply the result of a coordinate transformation
from the flat space metric can be seen by writing the metric in the
form\cite{ref:Millman_Parker}
\[
 g_{\mu\nu}=
 \left( \begin{array}{lll}{\displaystyle
\frac{\partial x^1}{\partial a^1}}& {\displaystyle\frac{\partial
x^2}{\partial
a^1}} & {\displaystyle\frac{\partial x^3}{\partial a^1}}\\[15pt]
 {\displaystyle\frac{\partial x^1}{\partial a^2}}& {\displaystyle\frac{\partial x^2}{\partial a^2}} &
{\displaystyle\frac{\partial x^3}{\partial a^2}}\\[15pt]
{\displaystyle\frac{\partial x^1}{\partial a^3}}&
{\displaystyle\frac{\partial x^2}{\partial a^3}} &
{\displaystyle\frac{\partial x^3}{\partial a^3}}
\end{array}
 \right)
 \left(\begin{array}{lll}
 {\displaystyle 1}& {\displaystyle 0} & {\displaystyle 0}\\[15pt]
 {\displaystyle 0}& {\displaystyle 1}& {\displaystyle 0}\\[15 pt]
 {\displaystyle 0}& {\displaystyle 0} & {\displaystyle 1}
 \end{array}
 \right)
 \left(\begin{array}{lll}
{\displaystyle \frac{\partial x^1}{\partial a^1}}&
{\displaystyle\frac{\partial x^1}{\partial
a^2}} & {\displaystyle \frac{\partial x^1}{\partial a^3}}\\[15pt]
 {\displaystyle\frac{\partial x^2}{\partial a^1}}& {\displaystyle\frac{\partial x^2}{\partial a^2}} &
{\displaystyle\frac{\partial x^2}{\partial a^3}}\\[15pt]
{\displaystyle\frac{\partial x^3}{\partial a^1}}&
{\displaystyle\frac{\partial x^3}{\partial a^2}} &
{\displaystyle\frac{\partial x^3}{\partial a^3}}
\end{array}
\right)
\]
\[
=J^TIJ
\]
where
\[
\frac{\partial x^\mu}{\partial a^\nu}=\delta_{\mu\nu}+\frac{\partial
u^\mu}{\partial a^\nu}.
\]
and $J$ is the Jacobian of the transformation. Later in section
\ref{sec:fourier_transform} we will show that the metric for the
Fourier modes of our system is not a simple coordinate
transformation.

The inverse matrix $(g^{ik})=(g_{ik})^{-1}$ can be obtained by
direct inversion of Equation~(\ref{eq:metric_tensor}) which would
yield components of $g^{ik}$ in terms of derivatives of $u^{i}$ with
respect to the internal coordinates $a^i$.  Later it will be useful
however to have the inverse metric stated in terms of derivatives of
$u^i$ with respect to the external coordinates $x^i$.

To accomplish this we write the inverse metric as
$(g^{ik})=(J^{-1})(J^{-1})^T$ where
\begin{equation}
J^{-1}=\left(\begin{array}{lll} {\displaystyle\frac{\partial
a^1}{\partial x^1}}& {\displaystyle\frac{\partial a^1}{\partial
x^2}} & {\displaystyle\frac{\partial a^1}{\partial x^3}}\\[15pt]
 {\displaystyle\frac{\partial a^2}{\partial x^1}}& {\displaystyle\frac{\partial a^2}{\partial x^2}} &
{\displaystyle\frac{\partial a^2}{\partial x^3}}\\[15pt]
{\displaystyle\frac{\partial a^3}{\partial x^1}}&
{\displaystyle\frac{\partial a^3}{\partial x^2}} &
{\displaystyle\frac{\partial a^3}{\partial x^3}}
\end{array}
\right)
\end{equation}
 This yields for the inverse metric
 \begin{eqnarray}
 \label{eq:inverse_metric}
 g^{ik}&=&\delta_{ik}-\frac{\partial
u^i}{\partial x^k}-\frac{\partial u^k}{\partial x^i}+\sum_l
\frac{\partial u^l}{\partial x^i}\frac{\partial u^l}{\partial x^k}\\
&=&\delta_{ik}+2\epsilon_{ik}\nonumber
 \end{eqnarray}
and $\epsilon_{ik}$ is the strain tensor in fixed space coordinates.
We see that the metric components can be written in terms of either
sets of coordinates, internal or fixed space.

\section{Equations of Motion}
\label{sec:EOM}
 In the following we will use the notation
\[
u_{\mu\nu}=\frac{\partial u^\mu}{\partial x^\nu}
\]
 and therefore the strain tensor is
 \[
 \epsilon_{\mu\nu}=\frac{1}{2}\left(-u_{\mu\nu}-u_{\nu\mu}+\sum_\beta
 u_{\beta \mu}u_{\beta\nu}\right).
 \]

We work in the fixed space coordinates and take the strain energy as
the lagrangian density of our system. This approach leads to the
usual equations of equilibrium in elasticity
theory\cite{ref:Love,ref:Novozhilov}. The strain energy is quadratic
in the strain tensor $\epsilon^{\mu\nu}$ and can be written as
\[
E=\sum_{\mu \nu\alpha\rho} C_{\mu \nu\alpha\rho}\, \epsilon_{\mu\nu}
\epsilon_{\alpha\rho}
\]

The quantities $C_{\mu \nu\alpha\rho}$ are known as the elastic
stiffness constants of the material\cite{ref:Sokolnikoff}.  For an
isotropic space most of the coefficients are zero and in $3$
dimensions, the lagrangian density reduces to
\begin{equation}
\label{eq:lagrangian_3D}
 L=(\lambda +
2\mu)\left[\epsilon_{11}^2+\epsilon_{22}^2+\epsilon_{33}^2\right] +
2 \lambda \left[\epsilon_{11} \epsilon_{22}+ \epsilon_{11}
\epsilon_{33} + \epsilon_{22}\epsilon_{33}\right] + 4\mu
\left[\epsilon_{12}^2 + \epsilon_{13}^2 + \epsilon_{23}^2\right]
\end{equation}
where $\lambda$ and $\mu$ are known as Lam\'e
constants\cite{ref:Sokolnikoff}.

The usual Lagrange equations,
\[
\sum_\nu\frac{d}{dx^\nu}\left(\frac{\partial L}{\partial u_{\rho
\nu}}\right) - \frac{\partial L}{\partial u^\rho}=0,
\]
apply with each component of the displacement vector treated as an
independent field variable. Since our Lagrangian contains no terms
in the field $u^\rho$, Lagrange's equations reduce to
\[
\sum_\nu\frac{d}{dx^\nu}\left(\frac{\partial L}{\partial u_{\rho
\nu}}\right)=0.
\]

which we shall denote
\[
V^\rho\equiv\sum_\nu\frac{d}{dx^\nu}\left(\frac{\partial L}{\partial
u_{\rho \nu}}\right)=0
\]

Using the above form of the Lagrangian one can write
\begin{eqnarray}
\label{eq:lagrangian_matrix} \frac{\partial L}{\partial u_{\rho
\nu}}&=&2\lambda(\sum_\alpha\epsilon_{\alpha\alpha})(\delta_{\rho\nu}-u_{\rho\nu})+
4\mu\sum_\alpha\epsilon_{\alpha\nu}(\delta_{\rho\alpha}-u_{\rho\alpha})\nonumber\\
&=& -2\lambda\sigma-2\mu(u_{\rho\nu}+u_{\nu\rho})+\lambda
\left[\sum_{\alpha}\left(u_{\alpha 1}^2+u_{\alpha
2}^2\right)\delta_{\rho\nu} - 2\left(\epsilon_{11}+
\epsilon_{22}\right)u_{\rho\nu}\right]\\
 & & \mbox{} \hspace{1.5in}
+2\mu\left[\sum_\alpha\sum_\beta
u_{\beta\alpha}u_{\beta\nu}\delta_{\rho\alpha} -
2\sum_\alpha\epsilon_{\alpha\nu}u_{\rho\alpha}\right]\nonumber
\end{eqnarray}
where the divergence of the displacement field is
$\sigma\equiv(u_{11}+u_{22}+u_{33})$.

 The first two terms in
Equation~(\ref{eq:lagrangian_matrix}) are first order in the
components $u_{\mu\nu}$ while the last terms (in square brackets)
are second order and higher in $u_{\mu\nu}$.

 Let us denote by $E_{\rho\nu}$ the last two bracketed terms in
 Equation~(\ref{eq:lagrangian_matrix}). This
allows us to write
\begin{equation}
\label{eq:lagrangian_matrix0}
V^\rho=-2\lambda\frac{\partial\sigma}{\partial x^\rho}
-2\mu\sum_\nu\frac{\partial}{\partial
x^\nu}(u_{\rho\nu}+u_{\nu\rho})+\sum_\nu\frac{\partial
E_{\rho\nu}}{\partial x^\nu}=0.
\end{equation}

The quantity $ V^\rho$ is a vector and every term in
Equation~(\ref{eq:lagrangian_matrix0}) transforms as a vector.  As
such the quantity
\[
E_\rho\equiv\sum_\nu\frac{\partial E_{\rho\nu}}{\partial x^\nu}
\]
can always be written as the gradient of a scalar plus the curl of a
vector
\[
\vec{E}=\nabla\alpha+\nabla\times\chi.
\]

This allows us to write
\begin{equation}
\label{eq:lagrangian_vector}
V^\rho=-2\lambda\frac{\partial\sigma}{\partial
x^\rho}-2\mu\left(\nabla^2u^{\rho}+\frac{\partial\sigma}{\partial
x^\rho}\right)+\frac{\partial\alpha}{\partial
x^\rho}+(\nabla\times\chi)^\rho
\end{equation}
or
\begin{equation}
\label{eq:laplaces_equation} \nabla^2\phi=0,
\end{equation}
where
\[\phi=[(2\lambda+4\mu)\sigma-\alpha]
\]
and notation $(\nabla\times\chi)^\rho$ represents the $\rho$
component of the vector $(\nabla\times\chi)$.
 We see therefore that there exists a scalar quantity,
in the medium that obeys Laplace's equation. Given that $\alpha$ is
second order in the strain quantities this implies that $\phi$
reduces to the divergence of the displacement field, $\sigma$, in
the infinitesimal strain approximation.
\subsection{Small Strain Approximation}
\label{sec:small_strain} In the usual treatment of elasticity theory
the strain components $u_\mu$ and their derivatives $u_{\mu\nu}$ are
taken to be infinitesimal. In this approximation, only the first
order terms in any equation are kept. We will not make this
assumption in this paper. Where necessary we will adopt a small
strain approximation where the strain components are small but not
infinitesimal.  In using this small strain approximation we will
always keep our equations of motion to at least order one order of
magnitude higher in the strain components than the infinitesimal
approximation.

\subsection{Internal Coordinates}
In sections \ref{sec:internal_coordinates} and
\ref{sec:interacting_mass}, we will need to translate the equations
of motion from the fixed space coordinates to the internal
coordinates.  For clarity and to adopt a more consistent convention,
in the remainder of this text we change notation slightly and write
the internal coordinates not as $a^i$ but as $x'^i$ and the fixed
space coordinates will continue to be unprimed and denoted $x^i$.
Now using $u^i=x^i-x'^i$ we can write

\begin{eqnarray}
\label{eq:coordinate_change} \frac{\partial}{\partial
x^i}&=&\sum_j\frac{\partial x'^j}{\partial
x^i}\frac{\partial}{\partial x'^j} \nonumber \\
&=& \sum_j\left(\frac{\partial
x^j}{\partial x^i}-\frac{\partial u^j}{\partial x^i}\right)\frac{\partial}{\partial x'^j}\nonumber\\
&=& \sum_j\left(\delta_{ij}-\frac{\partial u^j}{\partial
x^i}\right)\frac{\partial}{\partial x'^j}\nonumber\\
&=& \frac{\partial}{\partial x'^i}-\sum_j\frac{\partial
u^j}{\partial x^i}\frac{\partial}{\partial x'^j}
\end{eqnarray}

Equation~(\ref{eq:coordinate_change}) relates derivatives in the
fixed space coordinates $x^i$ to derivatives in the material
coordinates $x'^i$. As mentioned earlier, in the standard treatment
of elastic solids the displacements $u^i$ as well as their
derivatives are assumed to be infinitesimal and so the second term
in Equation~(\ref{eq:coordinate_change}) is dropped and there is no
distinction made between the $x^i$ and the $x'^i$ coordinates.  In
this paper we will keep the nonlinear terms in
Equation~(\ref{eq:coordinate_change}) when changing coordinates.
Hence we will make a distinction between the two sets of coordinates
and this will be pivotal in the derivations to follow.
\subsection{Vector Transformation}
The partial derivatives in Equation~(\ref{eq:coordinate_change})
transform as covectors. Therefore from
Equation~(\ref{eq:coordinate_change}) we see that any covector
transforms as

\begin{equation}
\label{eq:3D_covector_transformation}
V_\mu=V'_\mu-\sum_j\frac{\partial u^j}{\partial x^i}V'_j
\end{equation}
 when going between the unprimed
Euclidean space and the primed coordinate system. In a similar
manner the components of a vector change as

\begin{eqnarray*}
dx'^i&=&\sum_j\frac{\partial
x'^i}{\partial x^j}dx^j \\
&=& \sum_j\left(\frac{\partial
x'^i}{\partial x^j}-\frac{\partial u^i}{\partial x^j}\right)dx^j\\
&=& \sum_j\left(\delta_{ij}+\frac{\partial u^i}{\partial
x^j}\right)dx^j
\end{eqnarray*}
or for any vector
\begin{equation}
\label{3D_vector_transformation} V^{\prime
i}=V^i-\sum_j\frac{\partial u^i}{\partial x^j}V^j
\end{equation} where the "up" and "down" notation is used to
distinguish components of a vector from those of a covector.  These
transformations will be useful in later sections.
\subsection{Decomposition of Strain}
\label{sec:strain_decomposition}

The transformation in Equation~(\ref{3D_vector_transformation}) has
the form $\vec{V}\rightarrow\vec{V}-\delta \vec{V}$ where

\[
\delta\vec{V}^\nu=\sum_\mu\frac{\partial u^\nu}{\partial
x^\mu}V^\mu.
\]
 This can be decomposed as\cite{ref:Sokolnikoff}:
\begin{eqnarray}
\label{eq:strain_decomposition}
 \delta
 V^\mu&=&\sum_\mu\left(\frac{u_{\mu\nu}+u_{\nu\mu}}{2}+\frac{u_{\mu\nu}-u_{\nu\mu}}{2}\right)V^\mu\nonumber\\
 &=&\sum_\mu(e_{\mu\nu}+\omega_{\mu\nu})V^\mu.
 \end{eqnarray}

 The quantity $\omega_{\mu\nu}=1/2(u_{\mu\nu}-u_{\nu\mu})$ represents a local rigid
 body motion of the medium\cite{ref:Sokolnikoff}. The quantity $e_{\mu\nu}=1/2(u_{\mu\nu}+u_{\nu\mu})$
 represents what is usually termed "pure deformation" and for sufficiently small
 $u_{\mu\nu}$ we have $e_{\mu\nu}\approx\epsilon_{\mu\nu}$ .  This decomposition will
 be useful later in identifying the electromagnetic field vector.

We will now demonstrate a new method for reducing Laplace's Equation
(\ref{eq:laplaces_equation}) to Dirac's equation and compare this
method to the traditional Dirac reduction.

\section{Cartan's Spinors}
\label{sec:Cartan}
 The concept of Spinors was introduced by Eli
Cartan in 1913\cite{ref:Cartan}. In Cartan's original formulation
spinors were motivated by studying isotropic vectors which are
vectors of zero length. In three dimensions the equation of an
isotropic vector is
\begin{equation}
\label{eq:isotropic_vector}
 (x^1)^2 + (x^2)^2 + (x^3)^2=0
\end{equation}
for complex quantities $x^i$.  A closed form solution to this
equation is realized as
\begin{equation}
\label{eq:Cartan_spinor_solution}
\begin{array}{lccr}
{\displaystyle x^1 =-2\xi_0\xi_1,\ } & {\displaystyle
x^2=i(\xi_0^2+\xi_1^2),}&\ \mathrm{and}\ & {\displaystyle
x^3=\xi_0^2-\xi_1^2}
\end{array}
\end{equation}
 where the two quantities $\xi_i$ are
\[
\begin{array}{lcr}
{\displaystyle\xi_0=\pm\sqrt{\frac{x^3-\imath x^2}{2}}}& \
\mathrm{and} \ & {\displaystyle \xi_1=\pm\sqrt{\frac{-x^3-\imath
x^2}{2}}}
\end{array}.
\]
The two component object $\xi=(\xi_0,\xi_1)$ is a
spinor\cite{ref:Cartan} and any equation of the form
(\ref{eq:isotropic_vector}) has a spinor solution.

In the following we use the notation
$\partial_\mu\equiv\partial/\partial x^\mu$ and Laplace's equation
is written
\[\left(\partial_1^2+\partial_2^2+\partial_3^2\right)\phi=0.
\]
This equation can be viewed as an isotropic vector in the following
way. The components of the vector are the partial derivative
operators $\partial/\partial x^\mu$ acting on the quantity $\phi$.
As long as the partial derivatives are restricted to acting on the
scalar field $\phi$ it has a spinor solution given by
\begin{equation}
\label{eq:spinor0}
 \hat{\xi}_0^2=\frac{1}{2}\left(\frac{\partial}{\partial
x^3}-i\frac{\partial}{\partial x^2}\right)=\frac{\partial}{\partial
z^0}
\end{equation}
and
\begin{equation}
\label{eq:spinor1}
\hat{\xi}_1^2=-\frac{1}{2}\left(\frac{\partial}{\partial
x^3}+i\frac{\partial}{\partial x^2}\right)=\frac{\partial}{\partial
z^1}
\end{equation}
where
\[
\begin{array}{lcr}
 {\displaystyle z^0=x^3+ix^2}& \ \mathrm{and}\ & {\displaystyle
 z^1=-x^3+ix^2}
\end{array}
\]and the "hat" notation indicates that the quantities
$\hat{\xi}$ are operators.  The equations
\[
\hat{\xi}_0^2=\frac{\partial}{\partial z^0}
\]
and
\[
\hat{\xi}_1^2=\frac{\partial}{\partial z^1}
\]
are equations of fractional derivatives of order $1/2$ denoted
$\hat{\xi}_0=D^{1/2}_{z^0}$ and $\hat{\xi}_1=D^{1/2}_{z^1}$.
Fractional derivatives have the property that\cite{ref:Miller_Ross}
\[
D^{1/2}_{z}D^{1/2}_{z}=\frac{\partial}{\partial z}
\]
 and various methods exist for writing closed form solutions for these
 operators\cite{ref:Miller_Ross}.
%\begin{equation}
%D^{\frac{1}{2}}_z \phi=\frac{1}{\Gamma
%\left(\frac{1}{2}\right)}\frac{\partial}{\partial z}\int^z_0
%(z-t)^{-\frac{1}{2}}\phi(t)dt
%\end{equation}
The exact form for these fractional derivatives however, is not
important here.  The important thing to note is that a solution to
Laplace's equation can be written in terms of spinors which are
fractional derivatives.
\subsection{Spinor Properties}

If we assume that the fractional derivatives $\hat{\xi}_0$ and
$\hat{\xi}_1$ commute then we also have
\begin{eqnarray*}
(\hat{\xi}_0\hat{\xi}_1)^2&=&\hat{\xi}_0\hat{\xi}_0\hat{\xi}_1\hat{\xi}_1 \nonumber \\
&=&\frac{\partial}{\partial z^0}\frac{\partial}{\partial z^1}\nonumber \\
&=&-\frac{1}{4}\left(\partial_3-
\imath\partial_2\right)
\left(\partial_3+ \imath\partial_2\right)\nonumber \\
&=&-\frac{1}{4}\left(\partial_2^2+
\partial_3^2\right)\nonumber \\
&=&\frac{1}{4}\partial_1^2\nonumber\\
\end {eqnarray*}

Using this result combined with Equations~(\ref{eq:spinor0}) and
(\ref{eq:spinor1}) we may write for the components of our vector
\begin{equation}
\label{eq:derivative1_solution}
 \frac{\partial}{\partial x_1}=
-2\hat{\xi}_0\hat{\xi}_1
\end{equation}
\begin{equation}
\label{eq:derivative2_solution} \frac{\partial}{\partial
x_2}=\imath(\hat{\xi}_0^2 + \hat{\xi}_1^2)
\end{equation}
\begin{equation}
\label{eq:derivative3_solution}
 \frac{\partial}{\partial x_3}=\hat{\xi}_0^2 - \hat{\xi}_1^2.
\end{equation}

Upon complex conjugation we observe that
\[
\begin{array}{lcr}
\left(\hat{\xi}_0^2\right)^*=-\hat{\xi}_1^2 &\mathrm{ and }
&\left(\hat{\xi}_1^2\right)^*=-\hat{\xi}_0^2.
\end{array}
\]
which implies
\begin{equation}
\label{eq:complex_conjugate_spinor_components}
\begin{array}{lcr}
 \hat{\xi}^*_0=\imath\hat{\xi}_1
&\mathrm{and} &\hat{\xi}^*_1=\imath\hat{\xi}_0
\end{array}.
\end{equation}
Under complex conjugation therefore our vector becomes\vspace{15pt}
\[
\vspace{15pt}
\begin{array}{lccr}
 {\displaystyle\left(\frac{\partial}{\partial x^1}\right)^*= -\frac{\partial}{\partial
 x^1},}& \ \ \ \
{\displaystyle\left(\frac{\partial}{\partial
x^2}\right)^*=\frac{\partial}{\partial x^2}}& \ \ \mathrm{and} \ \ \
\  &
 {\displaystyle\left(\frac{\partial}{\partial
x^3}\right)^*=\frac{\partial}{\partial x^3}}.
\end{array}
\]
 We see
therefore that our solution is consistent only for an elastic medium
embedded in a pseudo-Euclidean space in which the vector
$\partial/\partial x^1$ is pure imaginary. If we write
\[
\frac{\partial}{\partial x^1}=\imath\frac{\partial}{\partial t}
\]
then Laplace's equation becomes the wave equation
\[\left(-\frac{\partial^2}{\partial t^2}+
\vec{\nabla}^2\right)\phi=0
\]
We will continue to work with $dx$ rather than $dt$ to avoid
carrying the minus sign in the computations.

\subsection{Matrix Form}
It can be readily verified that our spinors satisfy the following
equations
\begin{eqnarray*}
\left[\hat{\xi}_0 \frac{\partial}{\partial x^1}+ \hat{\xi}_1
\left(\frac{\partial}{\partial x^3}-i\frac{\partial}{\partial
x^2}\right)\right]\phi=0\\
 \left[\hat{\xi}_0\left(\frac{\partial}{\partial x^3} +
i\frac{\partial}{\partial
x^2}\right)-\hat{\xi}_1\frac{\partial}{\partial x^1}\right]\phi=0
\end{eqnarray*}
and in matrix form
\begin{equation}
\label{eq:dirac_matrix}
 \left(
  \begin{array}{lr}
{\displaystyle\frac{\partial}{\partial x^1}} &
{\displaystyle\frac{\partial}{\partial
x^3}-i\frac{\partial}{\partial
 x^2}} \\[15pt]
{\displaystyle\frac{\partial}{\partial
x^3}+i\frac{\partial}{\partial
 x^2}} & {\displaystyle -\frac{\partial}{\partial x^1}}
  \end{array}
   \right)
 \left(\begin {array}{c}
 {\displaystyle\hat{\xi}_0 }\\[20pt]
  {\displaystyle\hat{\xi}_1}
  \end{array}
  \right)  \phi=0
\end{equation}

The matrix
\[
 X=\left(\begin{array}{lr}
 {\displaystyle\frac{\partial}{\partial x^1}} & {\displaystyle\frac{\partial}{\partial x^3}-i\frac{\partial}{\partial
 x^2}} \\[15pt]
{\displaystyle\frac{\partial}{\partial
x^3}+i\frac{\partial}{\partial
 x^2}} & {\displaystyle -\frac{\partial}{\partial x^1}}
 \end{array}
 \right)
 \]
 is equal to the dot product of the vector $\partial_\mu\equiv\partial/\partial x^\mu$ with the pauli spin matrices
 \[
 X=\frac{\partial}{\partial x^1}\gamma^1 + \frac{\partial}{\partial
 x^2}\gamma^2 + \frac{\partial}{\partial x^3}\gamma^3
 \]
 where
 \[
 \begin{array}{ccc}
 \gamma^1=\left(\begin{array}{ll}
 1 & 0\\
 0 & -1
 \end{array}
 \right),&
\gamma^2=\left(\begin{array}{ll}
 0 & -i\\
 i & 0
 \end{array}
  \right),&
 \gamma^3=\left(\begin{array}{ll}
 0 & 1\\
 1 & 0
 \end{array}
 \right)
\end{array}
 \] are the Pauli matrices which satisfy the
 anticommutation relations
 \begin{equation}
\label{eq:anti_comm}
\{\tilde{\gamma}^{\mu},\tilde{\gamma}^{\nu}\}=2I \delta_{\mu\nu}.
\end{equation} where $I$ is the identity matrix.

Equation~(\ref{eq:dirac_matrix}) can be written
\begin{equation}
\label{eq:dirac_unstrained}
\sum_{\mu=1}^3\partial_\mu\gamma^\mu\xi=0.
\end{equation}
where we have used the notation $\xi\equiv \hat{\xi}\phi$. This
equation has the form of Dirac's equation in $3$ dimensions for a
noninteracting massless field $\xi$.

\subsection{Relation to the Dirac Decomposition}
The fact that Laplace' equation and Dirac's equation are related is
not new.  However the decomposition used here is not the same as
that used by Dirac. The usual method of connecting the second order
Laplace Equation to the first order Dirac equation is to operate on
Equation~(\ref{eq:dirac_unstrained}) from the left with
$\sum_{\nu=1}^3\gamma^{\nu}\partial_\nu$ giving

\begin{eqnarray}
\label{eq:dirac_to_laplace}
0&=&\sum_{\mu,\nu=1}^3\gamma^{\nu}\gamma^{\mu}\partial_\nu \partial_\mu\Psi(x)\nonumber\\
&=&\sum_{\mu,\nu=1}^3\frac{1}{2}\left(\gamma^{\nu}\gamma^{\mu}+\gamma^{\mu}\gamma^{\nu}\right)
\partial_\nu\partial_\mu\Psi(x) \nonumber \\
&=&\left(\partial_1^2+\partial_2^2+\partial_3^2\right)\Psi(x)
\end{eqnarray}
where $\Psi=(\alpha_1,\alpha_2)$ is a two component spinor and
Equation~(\ref{eq:anti_comm}) has been used in the last step.

This shows that Dirac's equation does in fact imply Laplace's
equation. The important thing to note about
Equation~(\ref{eq:dirac_to_laplace}) however, is that the three
dimensional Dirac's equation implies not one Laplace equation but
two in the sense that each component of the spinor $\Psi$ satisfies
this equation. Explicitly stated,
Equation~(\ref{eq:dirac_to_laplace}) reads
\[
\left(\begin{array}{cc}
 \partial_1^2+\partial_2^2+\partial_3^2 & 0\\
 0 & \partial_1^2+\partial_2^2+\partial_3^2
 \end{array}
 \right)\left(\begin{array}{c}
 \alpha_1\\
 \alpha_2\end{array}\right) =0
 \] for the independent scalars $\alpha_1, \alpha_2$.

Conversely, if one starts with Laplace's equation and tries to
recover Dirac's equation, it is necessary to start with two
independent scalars each independently satisfying Laplace's
equation. In other words, using the usual methods, it is not
possible to take a single scalar field that satisfies Laplace's
equation and recover Dirac's equation for a two component spinor.

What has been demonstrated in the preceding sections is that
starting with only one scalar quantity satisfying Laplace's equation
Dirac's equation for a two component spinor may be derived.
Furthermore any medium (such as an elastic solid) that has a single
scalar that satisfies Laplace's equation must have a spinor that
satisfies Dirac's equation and such a derivation necessitates the
use of fractional derivatives.

\section{Transformation to Internal Coordinates}
\label{sec:internal_coordinates} In section
\ref{sec:fourier_transform} we will take the $x'^3$ coordinate
to be periodic and we will derive equations for the Fourier
components of our fields.  Since the elastic solid is assumed to be
periodic in the internal coordinates we need to translate our
equations of motion from fixed space coordinates to internal
coordinates. Using Equation~(\ref{eq:coordinate_change}) we can
rewrite Equation~(\ref{eq:dirac_unstrained}), as

\begin{equation}
\label{eq:dirac_before_FT}
\sum_{\mu=1}^3\gamma^\mu\left(\partial_\mu'-\sum_\nu\frac{\partial
u^\nu}{\partial x^\mu} \partial_\nu'\right)\xi=0
\end{equation}
or
\[
\sum_{\mu=1}^3\gamma^{\prime\mu}\partial'_\mu\xi=0
\]

where $\partial'_\mu=\partial/\partial x'_\mu$ and
$\gamma^{\prime\mu}$ is given by
\begin{equation}
\label{eq:modified_gamma_matrices}
\gamma^{\prime\mu}=\gamma^\mu-\sum_{\alpha=1}^3
u_{\mu\alpha}\gamma^\alpha.
\end{equation} By Equation~(\ref{3D_vector_transformation}) these are
the gamma matrices expressed in the primed coordinate system.

The anticommutator of these matrices is
\begin{eqnarray*}
\{\gamma^{\prime\mu},\gamma^{\prime\nu}\}&=&
\{\gamma^\mu-\sum_\alpha u_{\mu\alpha}\gamma^\alpha,\gamma^\nu-\sum_\beta u_{\nu\beta}\gamma^\beta\}\\
&=&\{\gamma^\mu,\gamma^\nu\}-\sum_\beta
u_{\nu\beta}\{\gamma^\mu,\gamma^\beta\} - \sum_\alpha
u_{\mu\alpha}\{\gamma^\alpha,\gamma^\nu\}+
\sum_{\alpha\beta}u_{\mu\alpha}u_{\nu\beta}\{
\gamma^\alpha,\gamma^\beta\}\\
&=&2I\left(\delta_{\mu\nu}-\sum_\beta u_{\nu\beta}\delta_{\mu\beta}-
\sum_\alpha u_{\mu\alpha}\delta_{\alpha\nu}+ \sum_\alpha\sum_\beta
u_{\mu\alpha}u_{\nu_\beta} \delta_{\alpha\beta}\right)\\
&=&2I\left(\delta_{\mu\nu}-u_{\nu\mu}-u_{\mu\nu}+\sum_\alpha
u_{\mu\alpha}u_{\nu\alpha}\right)\\
&\equiv& 2Ig^{\mu\nu}
\end{eqnarray*}

This shows that the gamma matrices have the form of the usual
dirac's matrices in a curved space\cite{ref:Brill_Wheeler}.  To
further develop the form of Equation~(\ref{eq:dirac_before_FT}) we
have to transform the spinor properties of $\xi$.  As currently
written $\xi$ is a spinor with respect to the $x_i$ coordinates not
the $x'_i$ coordinates. To transform its spinor properties we use a
real similarity transformation\cite{ref:Brill_Wheeler} and write
$\xi=S\tilde{\xi}$ where $S$ is a similarity transformation that
takes our spinor in $x_\mu$ to a spinor in $x'_\mu$.

We then have
\[
\partial'_\mu\xi=(\partial'_\mu S)\tilde{\xi}+S\partial'_\mu\tilde{\xi}.
\]
Equation~(\ref{eq:dirac_before_FT}) then becomes
\[
\begin{array}{lcl}
 0&=&\gamma^{\prime\mu}
[S\partial'_\mu\tilde{\xi}+(\partial'_\mu S)\tilde{\xi}]\\
 \mbox{}
&=&\gamma^{\prime\mu}
S[\partial'_\mu\tilde{\xi}+S^{-1}(\partial'_\mu
S)\tilde{\xi}]\\
\mbox{} &=&S^{-1}\gamma^{\prime\mu}
S[\partial'_\mu\tilde{\xi}+S^{-1}(\partial'_\mu S)\tilde{\xi}]
\end{array}
\]
Using $(\partial'_\mu S^{-1}) S=-S^{-1}(\partial'_\mu S)$. This can
finally be written
\begin{equation}
\label{eq:dirac_curved_space} \tilde{\gamma}^\mu
[\partial'_\mu-\Gamma_\mu]\tilde{\xi}=0
\end{equation}
where $\Gamma_\mu=(\partial'_\mu S^{-1})S$ and
$\tilde{\gamma}^\mu=S^{-1}\gamma^{\prime\mu} S$.

Equation~(\ref{eq:dirac_curved_space}) has the form of the
Einstein-Dirac equation in 3 dimensions for a free particle of zero
mass. The quantity $\partial'_\mu-\Gamma_\mu$ is the covariant
derivative for an object with spin in a curved
space\cite{ref:Brill_Wheeler}. In order to make this identification,
the field $\Gamma_\mu$ must satisfy the additional
equation\cite{ref:Brill_Wheeler,ref:Brill_Cohen}
\[
\frac{\partial {\tilde{\gamma}^\mu}}{\partial
x^\nu}+\tilde{\gamma}^\beta\Gamma^{\prime\mu}_{\beta\nu}-
\Gamma_\nu\tilde{\gamma}^\mu+\tilde{\gamma}^\mu\Gamma_\nu=0
\]
where $\Gamma^{\prime\mu}_{\beta\nu}$ is the usual Christoffel
symbol. That this equation holds can be seen by considering the
equation
\[
\frac{\partial \gamma^\mu}{\partial x^\nu}=0
\]
true in the unprimed coordinate system.  But since the unprimed
coordinate system is Euclidean space, the Christoffel symbols are
identically zero. This allows us to write
\[
\frac{\partial \gamma^\mu}{\partial
x^\nu}+\gamma^\beta\Gamma^\mu_{\beta\nu}=0
\]
Since this is a tensor equation true in all frames, in the primed
coordinate system we can immediately write
\[
\partial'_\nu
\gamma^{\prime\mu}+\gamma^{\prime\beta}\Gamma^{\prime\mu}_{\beta\nu}=0
\]
Using $\gamma^{\prime\mu}=S\tilde{\gamma}^\mu S^{-1}$, we have

\[
 \partial'_\nu (S\tilde{\gamma}^\mu
S^{-1})+(S\tilde{\gamma}^\beta S^{-1})\Gamma'^\mu_{\beta\nu} =0
\]
or
\[(\partial'_\nu
S)\tilde{\gamma}^\mu S^{-1}+ S(\partial'_\nu \tilde{\gamma}^\mu)
S^{-1}+ S \tilde{\gamma}^\mu (\partial'_\nu
S^{-1})+(S\tilde{\gamma}^\beta S^{-1})\Gamma'^\mu_{\beta\nu}=0.
\]

Multiplying by $S^{-1}$ on the left and $S$ on the right yields
\[
S^{-1}(\partial'_\nu S)\tilde{\gamma}^\mu + (\partial'_\nu
\tilde{\gamma}^\mu) + \tilde{\gamma}^\mu (\partial'_\nu
S^{-1})S+\tilde{\gamma}^\beta \Gamma'^\mu_{\beta\nu} =0
\]

Finally, using $\Gamma_\nu=(\partial'_\nu S^{-1})S$ and again noting
that $\partial'_\nu S^{-1}S=-S^{-1}\partial'_\nu S$ we have,
\begin{equation}
\label{eq:aux_3D}
\tilde{\gamma}^\mu\Gamma_\nu-\Gamma_\nu\tilde{\gamma}^\mu
+\left(\partial'_\nu \tilde{\gamma}^\mu  +\tilde{\gamma}^\beta
\Gamma'^\mu_{\beta\nu}\right)=0
\end{equation}

We have just demonstrated that in the internal coordinates, the
equations of motion of an elastic medium have the same form as the
free-field Einstein Dirac equation for a massless particle in three
dimensions. Of course this form is trivial in the sense that it is
due solely to a coordinate change and can be removed by simply
changing back to the unprimed coordinates. In the next section we
will show that if one of the dimensions in our problem is periodic,
then the equations of motion for the fourier modes is not trivial.
Furthermore the introduction of the fourier components will generate
extra terms in Equation~(\ref{eq:dirac_before_FT}) that implies a
series of equations relevant for particles with mass coupled to
fields that can be associated with electromagnetism and classical
gravity.

\section{Interacting particles with mass}
\label{sec:interacting_mass}
 In this section we again consider a
three dimensional elastic solid but we take the third dimension to
be compact with the topology of a circle. All field variables then
become periodic functions of $x'^3$ and can be Fourier transformed.
The act of Fourier transforming the equations of motion will
effectively reduce the dimensionality of our problem from three
dimensions ($x'^1$,$x'^2$,$x'^3$) to two dimensions ($x'^1$,$x'^2$).
This dimensional reduction will in some cases create two of the same
type of objects (for instance a two dimensional metric and a three
dimensional metric). We will need to take care to distinguish
between the two dimensional and three dimensional quantities and use
explicit limits in most summations.

Throughout the derivation we will assume the small strain
approximation of section (\ref{sec:small_strain}).  In doing so, we
will keep the equations of motion of the system to one order of
magnitude higher in the strain components than the infinitesimal
results. For the Dirac equation the infinitesimal result would yield
$\gamma^\mu \partial_\mu \Psi=0$ which is a linear operator that is
zero'th order in $u_{\mu\nu}$, acting on $\Psi$. In our derivation
extra terms will appear in this equation of motion and we will keep
the linear operator to first order in $u_{\mu\nu}$.

\subsection{Fourier Transform}
\label{sec:fourier_transform} In preparation for Fourier
Transforming we isolate the terms involving $x'^3$ and rewrite
Equation~(\ref{eq:dirac_before_FT}) as,

\[
 \sum_{\mu=1}^2\gamma^\mu
\left(\partial_\mu'-\sum_{\nu=1}^2 u_{\nu\mu} \partial_\nu' - u_{3\mu}
\partial_3' \right)\xi
+ \gamma^3\left(\partial_3'-\sum_{\nu=1}^2 u_{\nu3} \partial_\nu' - u_{33}
\partial_3' \right)\xi=0
\]
Similar to Equation~(\ref{eq:modified_gamma_matrices}) we define
\[
\gamma'^\mu=\gamma^\mu-\sum_{\beta=1}^2 u_{\mu\beta}\gamma^\beta
\]
to obtain
\[
 \sum_{\mu=1}^2\gamma'^\mu
\left(\partial_\mu' - u_{3\mu}
\partial_3' \right)\xi -\left(\sum_{\beta=1}^2 \gamma^\beta
u_{\mu\beta}u_{3\mu}\partial'_3\right)\xi +
\gamma^3\left(\partial_3'-\sum_{\nu=1}^2 u_{\nu3}
\partial_\nu' - u_{33}
\partial_3' \right)\xi=0
\]
In keeping with the small strain approximation the second term in
parenthesis may be neglected as being second order in the small
strain quantities. We are left with
\[
 \sum_{\mu=1}^2\gamma'^\mu
\left(\partial_\mu' - u_{3\mu}
\partial_3' \right)\xi  +
\gamma^3\left(\partial_3'-\sum_{\nu=1}^2 u_{\nu3}
\partial_\nu' - u_{33}
\partial_3' \right)\xi=0
\]
 Next we transform the spinor
\[
\xi=S\tilde{\xi}
\]
Which yields
\begin{eqnarray*}
\sum_{\mu=1}^2\gamma'^\mu \left(S\left[\partial_\mu'-
u_{3\mu}\partial_3'\right] + \left[(\partial_\mu'S) - u_{3\mu}
(\partial_3'S)\right]\right)\tilde{\xi}
\hspace{.5in}\\
+ \gamma^3 \left(S\left[\partial_3'-\sum_{\nu=1}^2 u_{\nu3}
\partial_\nu' - u_{33}\partial_3'\right] +
(\partial_3'S)-\sum_{\nu=1}^2 u_{\nu3} (\partial_\nu'S) - u_{33}
(\partial_3'S)\right)\tilde{\xi}=0
\end{eqnarray*}
In the absence of deformation there is no distinction between primed
and unprimed coordinates and the similarity transformation therefore
reduces to the identity matrix as the displacement field $u_\mu$
goes to zero. This implies that $S$ has the form
\[
S=I+f(u_\mu)
\]
where $f(u_\mu)$ is a matrix that is a function of the strain
components  and $f\rightarrow0$ as the strain components go to zero.
This gives
\begin{eqnarray*}
u_{\alpha\beta}\partial'_\mu S&=&u_{\alpha\beta}\frac{\partial
f(u_\nu)}{\partial x'_\mu}\\
&=&u_{\alpha\beta}\sum_\nu\frac{\partial f}{\partial
u_\nu}\frac{\partial u_\nu}{\partial x'_\mu}\\
 & \sim &u_{\alpha\beta}u_{\nu\mu}\frac{\partial f}{\partial
u_\nu}
\end{eqnarray*}
which implies that these terms are second order in the strain
components and can be neglected.  We are left with
\begin{eqnarray*}
\sum_{\mu=1}^2\gamma'^\mu S\left(\partial_\mu' - u_{3\mu}\partial_3'
+ S^{-1}(\partial'_\mu S)\right)\tilde{\xi}
\hspace{1.5in}\\
+ \gamma^3 S\left(\partial_3'-\sum_{\nu=1}^2 u_{\nu3}
\partial_\nu' - u_{33}\partial_3' +
S^{-1}(\partial'_3 S)\right)\tilde{\xi}=0
\end{eqnarray*}

Now we multiply by $-\imath S^{-1}\gamma^3$ on the left to obtain
\begin{eqnarray}
\label{eq:dirac_curved_space_separated}
 \sum_{\mu=1}^2
\tilde{\gamma}^\mu\left(\partial_\mu'- u_{3\mu}\partial_3' +
S^{-1}(\partial'_\mu S)\right)\tilde{\xi}
\hspace{1.5in}\nonumber\\
- \imath\left(\partial_3'-\sum_{\nu=1}^2 u_{\nu3}
\partial_\nu' - u_{33}\partial_3' +
S^{-1}(\partial'_3 S)\right)\tilde{\xi}=0
\end{eqnarray}
where
\[\tilde{\gamma}^\mu =-\imath S^{-1}\gamma^3\gamma^\mu S.
\]
This translation to the internal coordinates has introduced several
fields into the equations of motion.  In addition to the spinor
field $\tilde{\xi}$ there are the fields $u_{\mu\nu}$ and
$S^{-1}(\partial'_\mu S)$.   We are now in a position to Fourier
transform these field quantities.  We first transform the fields
$u_{\mu\nu}$ and $S^{-1}(\partial_\mu S)$ in equation
(\ref{eq:dirac_curved_space_separated}) to obtain
\[
u_{\nu\mu}=\sum_ku_{\nu\mu,k}e^{ikx_3'}
\]
and
\[
S^{-1}(\partial_\mu S)=\left[S^{-1}(\partial_\mu S)\right]_k
e^{ik'x_3}
\]
 where $u_{\nu\mu,k}$ and $\left[S^{-1}(\partial_\mu S)\right]_k$ are
 the $k^{th}$ Fourier modes of the relevant quantities and $k=2\pi n/a$ with $a$ the length of the
circle formed by the elastic solid in the $x_3'$ direction and $n$
is an integer. Equation~(\ref{eq:dirac_curved_space_separated}) now
becomes,
\begin{eqnarray*}
\lefteqn{\sum_k e^{ikx_3'}\left[\sum_{\mu=1}^2
\tilde{\gamma}^\mu\left(\partial_\mu'\delta_{k,0} -
u_{3\mu,k}\partial_3' + \left[S^{-1}(\partial'_\mu
S)\right]_k\right)\tilde{\xi}
\right.}\hspace{0in}\\
 & & \left.\mbox{}
- \imath\left(\partial_3'\delta_{k,0}-\sum_{\nu=1}^2 u_{\nu3,k}
\partial_\nu' - u_{33,k}\partial_3' +
\left[S^{-1}(\partial'_3 S)\right]_k\right)\tilde{\xi}\right]=0
\end{eqnarray*}

Next we transform the spinor (remembering that it is periodic in
$4\pi$)
\[
\tilde{\xi}=\sum_q \tilde{\xi}_q e^{i(q/2)x_3'}
\]
and the gamma matrices
\[
\tilde{\gamma}^\mu=\sum_l\tilde{\gamma}^\mu_l e^{\imath l x_3'}
\]
with $q=2\pi j/a$, $l=2\pi j'/a$ and $j,j'$ integers. This yields,
\begin{eqnarray*}
\lefteqn{\sum_{q,k,l} e^{ix_3'(l+k+q/2)}\left[\sum_{\mu=1}^2
\tilde{\gamma}^\mu_l\left(\partial_\mu'\delta_{k,0} -
\imath\frac{q}{2}u_{3\mu,k} + \left[S^{-1}(\partial'_\mu
S)\right]_k\right)\tilde{\xi}_{q/2}
\right.}\hspace{1.5in}\\
 & & \left.\mbox{}
+ \delta_{l,0}\left(\frac{q}{2}\delta_{k,0}+\imath\sum_{\nu=1}^2
u_{\nu3,k}
\partial_\nu' - \frac{q}{2}u_{33,k}
-\imath\left[S^{-1}(\partial'_3
S)\right]_k\right)\tilde{\xi}_{q/2}\right]=0
\end{eqnarray*}

This equation is true for each distinct value of $l+k+q/2=m/2$ with
$m=2\pi n/a$ and $n$ an integer. Writing $q=(m-2k-2l)$ yields
finally,
\begin{eqnarray}
\label{eq:dirac_eq_all_modes} \lefteqn{\sum_{lk}
\left[\sum_{\mu=1}^2
\tilde{\gamma}^\mu_l\left(\partial_\mu'\delta_{k,0} -
\imath\frac{(m-2k-2l)}{2}u_{3\mu,k} + \left[S^{-1}(\partial'_\mu
S)\right]_k\right)\tilde{\xi}_{(m-2k-2l)/2}
\right.}\hspace{0in}\\
 & & \left.\mbox{}
+
\delta_{l,0}\left(\frac{(m-2k-2l)}{2}\delta_{k,0}-\imath\sum_{\nu=1}^2
u_{\nu3,k}
\partial_\nu' - \frac{(m-2k-2l)}{2}u_{33,k} -
\imath\left[S^{-1}(\partial'_3
S)\right]_k\right)\tilde{\xi}_{(m-2k-2l)/2}\right]=0\nonumber
\end{eqnarray}

 This is an infinite series
of equations describing the dynamics of the fields $\xi_{m/2}$.

\subsection{Spectrum of Lowest modes}
\label{sec:lowest_modes}
 In quantum mechanical systems a low energy
approximation can often be made in which only the lowest Fourier
modes of a system are present at low energies (or low temperatures).
For instance, in condensed matter systems the mean number of phonons
of wavevector $k$ present in a system at temperature $T$ is
\begin{equation}
\label{eq:occupation_n}
 n_k=\frac{1}{e^{\frac{\hbar\omega_k}{k_B
T}}-1}
\end{equation}
 where $\hbar$ is Planck's constant, $k_B$ is
Boltzmann's constant, T is temperature and $\omega_k$ for acoustic
phonons can be approximated by $\omega_k=ck$ with $c$ a constant.
The important aspect of Equation~(\ref{eq:occupation_n}) is that as
$T$ approaches $0$ only the $k=0$ mode is occupied.  So for spin
zero fields (such as $u_{\mu\nu,0}$) in the low energy limit in a
quantum system only the $k=0$ modes are present. In this paper we
are not considering a quantum mechanical system.  Nevertheless we
shall examine this "low energy" approximation but will not attempt
to give a rigorous justification, focusing instead on examining the
form of the equations of motion that result.

We therefore consider a theory in which only the lowest $k$ and $l$
modes are present and Equation~(\ref{eq:dirac_eq_all_modes}) reduces
to
\begin{eqnarray*}
\lefteqn{\sum_{\mu=1}^2
\tilde{\gamma}^\mu_0\left(\partial_\mu'-\sum_{\nu=1}^2 u_{\nu\mu,0}
\partial_\nu' - \imath\frac{m}{2}u_{3\mu,0} + \left[S^{-1}(\partial'_\mu
S)\right]_0\right)\tilde{\xi}_{m/2}
}\hspace{.5in}\\
 & & \mbox{}
+ \left(\frac{m}{2}+\imath\sum_{\nu=1}^2 u_{\nu3,0}
\partial_\nu' - \frac{m}{2}u_{33,0} -
\imath\left[S^{-1}(\partial'_3 S)\right]_0\right)\tilde{\xi}_{m/2}=0
\end{eqnarray*}
true for each value of $m$.

We are now in a position to examine in detail the form of the
equations for the spinor fields $\xi_{m/2}$.

\subsubsection{$m_{1}$ mode}
\label{sec:mode1}  We use the notation $m_q=2\pi q/a$ corresponding
to the equation of motion of the fields $\xi_{q/2}$. For clarity we
focus on the equation of motion of the $q=1$ field which yields
\begin{eqnarray}
\label{eq:mode1}
 \lefteqn{\sum_{\mu=1}^2 \tilde{\gamma}^\mu_0\left(\partial_\mu'
  - \imath\frac{m_1}{2}u_{3\mu,0} + \left[S^{-1}(\partial'_\mu
S)\right]_0\right)\tilde{\xi}_{1/2}
}\hspace{.5in}\\
 & & \mbox{}
+ \left(\frac{m_1}{2}+\imath\sum_{\nu=1}^2 u_{\nu3,0}
\partial_\nu' - \frac{m_1}{2}u_{33,0} -\imath
\left[S^{-1}(\partial'_3
S)\right]_0\right)\tilde{\xi}_{1/2}=0\nonumber
\end{eqnarray}

Let us focus on the first three terms in Equation~(\ref{eq:mode1})
\begin{equation}
\label{eq:mode1_b} \sum_{\mu=1}^2 \tilde{\gamma}^{\mu}_0
\left(\partial_\mu' - \imath\frac{m_1}{2}u_{3\mu,0} +
\left[S^{-1}(\partial'_\mu S)\right]_0\right)\tilde{\xi}_{1/2}.
\end{equation}

This portion of the equation of motion is very suggestive when
compared with Equation~(\ref{eq:full_dirac}) and suggests we
associate $\tilde{\gamma}^\mu$ with the $2$ dimensional gamma
matrices and that we associate $u_{3\mu,0}$ and
$\left[S^{-1}(\partial'_\mu S)\right]_0$ with the electromagnetic
vector potential and the spin connection.  In order to make these
identifications we must show that they are consistent with Maxwell's
equations, and the auxiliary Equation~(\ref{eq:auxiliary}).  We will
now demonstrate that these definitions are consistent.
\subsection{Spin Connection}
\label{sec:spin_connection}
 We propose to make the following
definition for the 2D spin connection
\begin{equation}
\label{eq:2D_spin_connection} \Gamma_\mu=\left[(\partial_\mu S^{-1})
S\right]_0+\imath \left(\frac{m_1}{2}\right)\omega_{3\mu}
\end{equation}
where $\omega_{3\mu}=(1/2)(u_{3\mu}-u_{\mu 3})$.   It must now be
shown that this is consistent with
  Equation~(\ref{eq:auxiliary}).  A solution to Equation~(\ref{eq:auxiliary})
  is unique only up to an additive multiple of the unit matrix\cite{ref:Brill_Wheeler,ref:Fletcher}.
  It is sufficient therefore to
  consider only the first term on the right hand side of
  Equation~(\ref{eq:2D_spin_connection}).

First let us note that the gamma matrices satisfy the
anticommutation relations
\[
\left\{\tilde{\gamma}^\mu,\tilde{\gamma}^\nu\right\}=
2I\left(\delta_{\mu\nu} -
(u_{\mu\nu}+u_{\nu\mu})+\sum_{\beta=1}^2u_{\mu\beta}u_{\nu\beta}\right).
\]
Upon Fourier transforming the gamma matrices this becomes
\[
\sum_{k,k'}e^{\imath
x'_3(k+k')}\left\{\tilde{\gamma}^\mu_k,\tilde{\gamma}^\nu_{k'}\right\}=
\sum_{k,k'}e^{\imath x'_3(k+k')}
2I\left(\delta_{\mu\nu}\delta_{k,k'} -
(u_{\mu\nu,k}+u_{\nu\mu,k})\delta_{k',0}+\sum_{\beta=1}^2u_{\mu\beta,k}u_{\nu\beta,k'}\right).
\]and using the ansatz that only the $k,k'=0$ modes are present
yields
\begin{equation}
\left\{\tilde{\gamma}^\mu_0,\tilde{\gamma}^\nu_{0}\right\}=
2I\left(\delta_{\mu\nu} -
(u_{\mu\nu,0}-u_{\nu\mu,0})+\sum_{\beta=1}^2u_{\mu\beta,0}u_{\nu\beta,0}\right).
\end{equation}
 If we
insist that the metric of our system is equal to twice the
anticommutator of the gamma matrices then we are led to define

\begin{equation}
\label{eq:metric_2D}
2I~g^{\mu\nu}\equiv \left\{\tilde{\gamma}^{\mu},\tilde{\gamma}^{\nu}\right\}\\
\end{equation}
which becomes
\begin{eqnarray}
\label{eq:metric_2D}
2I g^{\mu\nu}&\rightarrow& \left\{\tilde{\gamma}^{\mu}_0,\tilde{\gamma}^{\nu}_0\right\}\nonumber\\
&=&\delta_{\mu\nu} -
(u_{\mu\nu,0}+u_{\nu\mu,0})+\sum_{\beta=1}^2u_{\mu\beta,0}u_{\nu\beta,0}
.
\end{eqnarray}
 This is the metric for our two dimensional subspace and it does
not have the form of a simple coordinate transformation on a flat
space metric like that of section \ref{sec:elasticity_theory}. In
other words there is no transformation involving the coordinates
$x'^1$ and $x'^2$ that will remove the Fourier transform in this
equation.

The auxiliary Equation~(\ref{eq:auxiliary}) can now be shown to hold
in a similar manner to the construction in three dimensions.  First
we note that the definition of the two dimensional metric
Equation~(\ref{eq:metric_2D}) can be obtained from a transformation
on $2D$ Euclidean space

\begin{eqnarray}
 g^{\mu\nu}&=&\delta_{\mu\nu} -
(u_{\mu\nu}+u_{\nu\mu})+\sum_{\beta=1}^2u_{\mu\beta}u_{\nu\beta}\\
 &=&\left(\begin{array}{ll}{\displaystyle
\frac{\partial x^1}{\partial x'^1}}& {\displaystyle\frac{\partial
x^2}{\partial
x'^1}}\\[15pt]
 {\displaystyle\frac{\partial x^1}{\partial x'^2}}
 & {\displaystyle\frac{\partial x^2}{\partial x'^2}}
\end{array}
 \right)
 \left(\begin{array}{ll}
 {\displaystyle 1}& {\displaystyle 0}\\[15pt]
 {\displaystyle 0}& {\displaystyle 1}
 \end{array}
 \right)
 \left(\begin{array}{lll}
{\displaystyle \frac{\partial x^1}{\partial x'^1}}&
{\displaystyle\frac{\partial x^1}{\partial x'^2}} \\[15pt]
 {\displaystyle\frac{\partial x^2}{\partial x'^1}}& {\displaystyle\frac{\partial x^2}{\partial x'^2}}

\end{array}
\right)
\end{eqnarray}
\[
=J^TIJ
\]
where
\begin{equation}
\label{eq:2D_covector_transfomation}
 \frac{\partial x^\mu}{\partial
x'^\nu}=\delta_{\mu\nu}-\sum_{\nu=1}^2\frac{\partial u_\mu}{\partial
x'^\nu}.
\end{equation}
 In other
words the two dimensional metric in Equation~(\ref{eq:metric_2D}) is
the Fourier Transform of a matrix obtained by a coordinate
transformation from two dimensional Euclidean space. Given this
transformation we note, similar to
Equation~(\ref{3D_vector_transformation}), the components of any
vector $V^{\prime\mu}$ transform in the same manner as the
differential element $dx'^\mu$ or
\[
V^{\prime\mu}=V^\mu-\sum_{\beta=1}^2 u_{\mu\beta}V^\beta.
\]

We now start with the following tensor equation valid in the
Euclidean frame
\[
\frac{\partial \gamma^\mu}{\partial
x^\nu}=\partial_\nu\gamma^\mu+\sum_{\beta=1}^2\gamma^\beta\Gamma^\mu_{\beta\nu}=0
\]
which implies
\[
\partial'_\nu
\gamma^{\prime\mu}+\sum_{\beta=1}^2
\gamma^{\prime\beta}\Gamma^{\prime\mu}_{\beta\nu}=0
\]
or
\[
\partial'_\nu
(-\imath\gamma^3\gamma^{\prime\mu})+\sum_{\beta=1}^2
(-\imath\gamma^3\gamma^{\prime\beta})\Gamma^{\prime\mu}_{\beta\nu}=0
\]
valid in the primed coordinate system. We now use
$-\imath\gamma^3\gamma^{\prime\mu}=S\tilde{\gamma}^\mu S^{-1}$
 to obtain,
\[
-(\partial'_\nu S^{-1})S\tilde{\gamma}^\mu + \tilde{\gamma}^\mu
(\partial'_\nu S^{-1})S +\left(\partial'_\nu \tilde{\gamma}^\mu
+\sum_{\beta=1}^2\tilde{\gamma}^\beta
\Gamma^{\prime\mu}_{\beta\nu}\right)=0
\]

We now Fourier transform the fields $u_{\mu\nu}$ and $(\partial_\nu
S^{-1})S$ and use the ansatz that only the lowest fourier mode is
retained.  This gives
\begin{eqnarray*}
(\partial_\nu S^{-1})S&\rightarrow &\left[(\partial_\nu
S^{-1})S\right]_0\\
%u_{\mu\nu}&\rightarrow &u_{\mu\nu,0}\\
\tilde{\gamma}^{\mu}&\rightarrow&\tilde{\gamma}^{\mu}_0\\
g^{\mu\nu}&\rightarrow&\delta_{\mu\nu}-(u_{\mu\nu,0}+u_{\nu\mu,0})+\sum_{\beta=1}^2
u_{\beta\mu,0}u_{\beta\nu,0}
%\Gamma^\mu_{\beta\nu}&\rightarrow&\frac{1}{2}g^{\mu\alpha}\left(\frac{\partial
%g_{\alpha\beta}}{\partial x_\nu}+\frac{\partial
%g_{\alpha\nu}}{\partial x_\beta}- \frac{\partial
%g_{\beta\nu}}{\partial x_\alpha}\right)
\end{eqnarray*}
 This demonstrates that Equation~(\ref{eq:auxiliary}) is satisfied.

We next examine Maxwell's equations and the Einstein Field equations
by comparing our dimensional reduction to that used in Kaluza Klein
theory.

\section{Kaluza Klein Theory}
\label{sec:KK_theory}
 In Kaluza Klein Theories one starts with an $N$
dimensional system (usually $5$ dimensional) and uses dimensional
reduction to reduce the dimensionality of the system to $N-1$
dimensions.   The starting point is the assumption that the Ricci
tensor is identically zero in $N$ dimensions where the Ricci tensor
is defined as
\begin{equation}
\label{eq:ricci_tensor}
 R_{\alpha\beta}=\frac{\partial \Gamma^\rho_{\alpha\beta}}{\partial
 x^\rho} -\frac{\partial \Gamma^\rho_{\alpha\rho}}{\partial
 x^\beta}+ \Gamma^\sigma_{\alpha\beta}\Gamma^\rho_{\rho\sigma} -
  \Gamma^\sigma_{\alpha\rho}\Gamma^\rho_{\beta\sigma}
\end{equation}
and $\Gamma^\rho_{\alpha\beta}$ are the Christoffel symbols.

 The ansatz is then made that all field variables are
independent of the $N$th coordinate effectively reducing the
dimensionality of the problem to $N-1$ dimensions.  The key outcome
of this dimensional reduction is that the equation
$R_{\alpha\beta}=0$ in $N$ dimensions now becomes three sets of
equations in $N-1$ dimensions. Two of these three equations has the
form of Maxwell's equations and the Einstein Field equations.

In the next section we will compare our dimensional reduction with
the Kaluza Klein methods.  In doing so we will continue to use the
small strain approximation of Section~(\ref{sec:small_strain}).  Our
approach will be to keep the field equations to at least second
order in the strain quantities. From
Equation~(\ref{eq:ricci_tensor}) this implies that we need to keep
the Christoffel symbols to second order in $u_{\mu\nu}$.  Given the
form of the Christoffel symbols
\[
\Gamma^\rho_{\alpha\beta}=g^{\lambda\rho}\left(\frac{\partial
g_{\lambda\alpha}}{\partial x^\beta}+ \frac{\partial
g_{\lambda\beta}}{\partial x^\alpha}-\frac{\partial
g_{\alpha\beta}}{\partial x^\lambda}\right)
\]
we see that it is sufficient to keep the metric tensor
$g_{\alpha\beta}$ to second order in $u_{\alpha\beta}$ and the inverse
metric $g^{\alpha\beta}$ to first order in $u_{\alpha\beta}$.

\section{The Einstein Field and Maxwell's Equations}
\label{sec:EM_equation} In preparation for a comparison with Kaluza
Klein theory, we wish to invert the matrices defined in
Equation~(\ref{eq:inverse_metric}) and Equation~(\ref{eq:metric_2D})
keeping only terms that are to second order and lower in
$u_{\mu\nu}$.

In the following we will use the notation that $3$ dimensional
quantities are denoted with a carat (e.g. $\hat{g}_{\mu\nu}$) and
two dimensional quantities do not have a carat.  Our two dimensional
inverse metric from Equation~(\ref{eq:metric_2D}) is
\begin{eqnarray}
\label{eq:2D inverse_metric_form_2}
g^{\mu\nu}&=&\delta_{\mu\nu}+2\epsilon_{\mu\nu}\nonumber\\
&=&\left(\begin{array}{cc} {\displaystyle
1+2\epsilon_{11}}& {\displaystyle -2\epsilon_{12}}\\[15pt]
{\displaystyle -2\epsilon_{12}}& {\displaystyle 1+2\epsilon_{22}}
\end{array}\right)
\end{eqnarray}
where
\[\epsilon_{\mu\nu}=\frac{1}{2}\left(-u_{\mu\nu}-u_{\nu\mu}+\sum_{\beta=1}^2u_{\beta\nu}u_{\beta\mu}\right)
\]and $\mu,\nu$ take values $1$ or $2$.

The inverse of Equation~(\ref{eq:2D inverse_metric_form_2}) is
\[
g_{\mu\nu}=\frac{1}{D}\left(\begin{array}{cc} {\displaystyle
1+2\epsilon_{22}}& {\displaystyle -2\epsilon_{12}}\\[15pt]
{\displaystyle -2\epsilon_{12}}& {\displaystyle 1+2\epsilon_{11}}
\end{array}\right)
\]
The determinant of the inverse metric $g^{\mu\nu}$ is
\[
D=1+2(\epsilon_{11}+\epsilon_{22})+
4(\epsilon_{11}\epsilon_{22}-\epsilon_{12}^2)
\]

Using the expansion
\[
\frac{1}{1+x}=1-x+x^2+O(x^3)
\]
we have
\[
\frac{1}{D}=1-2(\epsilon_{11}+\epsilon_{22})+4(\epsilon_{11}\epsilon_{22}+
\epsilon_{12}^2+ \epsilon_{11}^2+\epsilon_{22}^2)+O(\epsilon^3)
\]

This gives the following form for the metric tensor correct to
second order in $u_{\mu\nu}$
\begin{equation}
\label{eq:2D_metric_tensor_second_order}
g_{\mu\nu}=\delta_{\mu\nu}-2\epsilon_{\mu\nu}+4\sum_{\alpha=1}^2\epsilon_{\alpha\mu}\epsilon_{\alpha\nu}
\end{equation}

In a similar manner the three dimensional metric in
Equation~(\ref{eq:inverse_metric}) can be inverted to yield
\begin{equation}
\label{eq:3D_metric_tensor_second_order}
\hat{g}_{\mu\nu}=\delta_{\mu\nu}-2\hat{\epsilon}_{\mu\nu}+
4\sum_{\alpha=1}^3\hat{\epsilon}_{\alpha\mu}\hat{\epsilon}_{\alpha\nu}
\end{equation}
with
\[
\hat{\epsilon}_{\mu\nu}=\frac{1}{2}\left(-u_{\mu\nu}-u_{\nu\mu}+\sum_{\beta=1}^3u_{\beta\nu}u_{\beta\mu}\right).
\] and $\mu,\nu$ taking on the values $1$ to $3$.
The combination of Equations~(\ref{eq:inverse_metric}),
(\ref{eq:metric_2D}), (\ref{eq:2D_metric_tensor_second_order}) and
(\ref{eq:3D_metric_tensor_second_order}) allows us to write the
three dimensional metric in terms of the two dimensional metric

\[
\hat{g}_{\alpha\beta}=\left(\begin{array}{cc} {\displaystyle
g_{\alpha\beta}+4\hat{\epsilon}_{3\alpha}\hat{\epsilon}_{3\beta}}\hspace{.1in}&
{\displaystyle -2\hat{\epsilon}_{\alpha 3}+4\sum_{\nu=1}^3\hat{\epsilon}_{\nu \alpha}\hat{\epsilon}_{\nu 3}}\\[15pt]
 {\displaystyle -2\hat{\epsilon}_{\beta 3}+4\sum_{\nu=1}^3\hat{\epsilon}_{\nu \beta}\hat{\epsilon}_{\nu 3}}
 & {\displaystyle 1-2\hat{\epsilon}_{3 3}+4\sum_{\nu=1}^3\hat{\epsilon}_{\nu 3}^2}
\end{array}
 \right)+O(u^3)
 \]
 and
\begin{equation}
\label{eq:2nd_order_metric}
 \hspace{.5in}
 \hat{g}^{\alpha\beta}=\left(\begin{array}{cc} {\displaystyle
g^{\alpha\beta}+u_{3\alpha}u_{3\beta}}\hspace{.1in}&
{\displaystyle -u_{\alpha 3}-u_{3\alpha}+\sum_{\nu=1}^3u_{\nu \alpha}u_{\nu 3}}\\[15pt]
 {\displaystyle -u_{\beta 3}-u_{3\beta}+\sum_{\nu=1}^3u_{\nu \beta}u_{\nu 3}}
 & {\displaystyle 1-2u_{33} + \sum_{\nu=1}^3 u_{\nu 3}^2}
\end{array}\right)+O(u^3)
\end{equation}

we can now compare this dimensional reduction to the usual Kaluza
Klein result. The Kaluza Klein dimensional reduction from $N$
dimensions to $N-1$ dimensions takes the form

\begin{equation}
\label{eq:KK_metric}
 \hat{g}(kk)_{\alpha\beta}=\left(\begin{array}{cc}
{\displaystyle
g_{\alpha\beta}-\Phi^2 A_\alpha}\hspace{.1in}& {\displaystyle-\Phi^2 A_\alpha}\\[15pt]
 {\displaystyle -\Phi^2A_\beta}
 & {\displaystyle-\Phi^2}
\end{array}
 \right)
 \hspace{.3in}
 \hat{g}(kk)^{\alpha\beta}=\left(\begin{array}{cc}
 {\displaystyle g^{\alpha\beta}}& {\displaystyle -A^\alpha}\\[15pt]
 {\displaystyle -A^\beta}& \hspace{.1in}{\displaystyle \left(-\Phi^2+A^\mu A_\mu\right)}
 \end{array}
 \right)
 \end{equation}
where $\vec{A}$ is the electromagnetic vector potential and $\Phi$
is a scalar. Comparison of Equations~(\ref{eq:2nd_order_metric}) and
(\ref{eq:KK_metric}) suggests that we make the following definitions
\begin{equation}
\label{eq:vector_potential}
A^\alpha=-\left(-u_{3\alpha}-u_{\alpha 3}+\sum_{\beta=1}^3u_{\beta\alpha}u_{\beta
3}\right)
\end{equation}
\begin{equation}
\label{eq:scalar_potential}
\Phi^2=-\left(1-2\hat{\epsilon}_{33}+4\sum_{\beta=1}^3\hat{\epsilon}_{\beta 3}^2\right)
\end{equation}
which implies
\begin{eqnarray}
\label{eq:vector_potential_lower}
A_\nu&=&\sum_{\mu=1}^2 g_{\mu\nu}A^\mu\nonumber\\
&=&\sum_{\mu=1}^2\left(\delta_{\mu\nu}-2\epsilon_{\mu\nu}+4\sum_{\beta=1}^2\epsilon_{\beta\mu}\epsilon_{\beta\nu}\right)
\left(u_{3\mu}+u_{\mu 3}-\sum_{\alpha=1}^3u_{\alpha 3}u_{\alpha\mu}\right)\nonumber\\
&=&\sum_{\mu=1}^2\left(\delta_{\mu\nu}-2\epsilon_{\mu\nu}+4\sum_{\beta=1}^2\epsilon_{\beta\mu}\epsilon_{\beta\nu}\right)
\left(-2\epsilon_{3\mu}-u_{3 3}u_{3\mu}\right)\\
&=&-2\epsilon_{3\nu}+4\sum_{\mu=1}^2\epsilon_{\mu\nu}\epsilon_{3\mu}-u_{33}u_{3\nu}+O(u^3)
\end{eqnarray}
and
\begin{eqnarray*}
-\left(\Phi^2\right)^{-1}&=&\frac{1}{1+\left(-2\hat{\epsilon}_{33}+
4\sum_{\beta=1}^3\hat{\epsilon}_{\beta 3}^2\right)}\\
&=&1+2\hat{\epsilon}_{33}- 4\sum_{\beta=1}^2\hat{\epsilon}_{\beta
3}^2
\end{eqnarray*}

This now allows us to put Equation~(\ref{eq:2nd_order_metric}) into
the same form as the Kaluza Klein decomposition,
\begin{equation}
\label{eq:metric_final}
 \hat{g}_{\alpha\beta}=\left(\begin{array}{cc}
{\displaystyle
g_{\alpha\beta}-\Phi^2 A_\alpha A_\beta}\hspace{.1in}& {\displaystyle-\Phi^2 A_\alpha}\\[15pt]
 {\displaystyle -\Phi^2A_\beta A_\alpha}
 & {\displaystyle-\Phi^2}
\end{array}
 \right)+O(u^3)
 \end{equation}
 \begin{equation}
 \hspace{.3in}
 \hat{g}^{\alpha\beta}=\left(\begin{array}{cc}
 {\displaystyle g^{\alpha\beta}}& {\displaystyle -A^\alpha}\\[15pt]
 {\displaystyle -A^\beta}& \hspace{.1in}{\displaystyle \left(-\Phi^{-2}+A^\mu A_\mu\right)}
 \end{array}
 \right)+O(u^2).
 \end{equation}
 The final
form is obtained by taking the Fourier transform and keeping only
the zero'th term of $u_{\mu\nu}$
\begin{eqnarray*}
A^\alpha &\rightarrow& \left(u_{3\alpha,0}+u_{\alpha
3,0}-\sum_{\beta=1}^3 u_{\beta\alpha,0}u_{\beta
3,0}\right)\\
g^{\mu\nu}&\rightarrow&\delta_{\mu\nu}-(u_{\mu\nu,0}+u_{\nu\mu,0})+\sum_{\beta=1}^2
u_{\beta\mu,0}u_{\beta\nu,0}\\
-\Phi^2&\rightarrow&\left(1-2\hat{\epsilon}_{33,0}+4\sum_{\beta=1}^3\hat{\epsilon}_{\beta
3,0}^2\right)
\end{eqnarray*}with
\[
\hat{\epsilon}_{\mu\nu,0}\equiv\frac{1}{2}\left(-u_{\mu\nu,0}-u_{\nu\mu,0}+
\sum_{\beta=1}^3u_{\beta\nu,0}u_{\beta\mu,0}\right)
\]
The Christoffel symbols and the Ricci scalar can now be written in
terms of the two dimensional metric, the electromagnetic vector
potential and the scalar field $\phi$. The details of this reduction
can be found in reference~{\cite{ref:Liu_Wesson} and here we simply
quote the results. Using Equation~(\ref{eq:metric_final}), the
Christoffel symbols are
\begin{eqnarray}
\hat{\Gamma}^\alpha_{\beta\gamma}&=&\Gamma^\alpha_{\beta\gamma}+
\frac{1}{2}\Phi^2\left(A_\beta F^\alpha_{\gamma}+ A_\gamma
F^\alpha_{\beta}\right) + \Phi\Phi^\alpha A_\beta A_\gamma \nonumber \\
\hat{\Gamma}^\alpha_{33}&=&\Phi \Phi^\alpha \nonumber\\
\hat{\Gamma}^\alpha_{\beta 3}&=&\frac{1}{2}\Phi^2F^\alpha_\beta +
\Phi\Phi^\alpha A_\beta\nonumber\\
\hat{\Gamma}^3_{3 3}&=&-\Phi\Phi^\gamma A_\gamma \nonumber\\
\hat{\Gamma}^3_{3\beta}&=&\frac{1}{2}\Phi^2 A^\gamma
F_{\lambda\beta}- \Phi\Phi^\lambda A_\lambda A_\beta
+\Phi^{-1}\Phi_\beta\nonumber\\
 \hat{\Gamma}^3_{\alpha\beta}&=&\frac{1}{2}\left(A_{\alpha;\beta}+ A_{\beta;\alpha}\right)- \frac{1}{2}\Phi^2
 A^\lambda\left(A_\alpha F_{\lambda\beta}+ A_\beta
F_{\lambda\alpha}\right)-A_\alpha A_\beta \Phi\Phi^\lambda
A_\lambda\nonumber\\
&&\hspace{1in}+\Phi^{-1}\left(A_\alpha\Phi_\beta+A_\beta\Phi_\alpha\right)
\end{eqnarray}
and the components of the Ricci tensor are,
\begin{eqnarray}
\hat{R}_{44}&=&\frac{1}{4}\Phi^4F^{\alpha\beta}F_{\alpha\beta}+\Phi\Phi^\alpha_{;\alpha}\nonumber\\
\hat{R}_{4\alpha}&=&\frac{1}{2}\Phi^2F^\lambda_{\alpha;\lambda}+
\frac{3}{2}\Phi\Phi^\lambda F_{\lambda\alpha}+
A_\alpha\left(\frac{1}{4}\Phi^4F^{\mu\nu}F_{\mu\nu}+\Phi\Phi^\mu_{;\mu}\right)\nonumber\\
\hat{R}_{\alpha\beta}&=&R_{\alpha\beta}+\frac{1}{2}\Phi^2F_{\alpha\lambda}F_\beta^\lambda-\Phi^{-1}\Phi_{\alpha;\beta}+
\frac{1}{2}A_\alpha\left(\Phi^2F^\gamma_{\beta;\gamma}+3\Phi\Phi^\gamma
F_{\gamma\beta}\right)\nonumber\\
&&\hspace{.5in}+\frac{1}{2}A_\beta\left(\Phi^2F^\gamma_{\alpha;\gamma}+3\Phi\Phi^\gamma
F_{\gamma\alpha}\right)+A_\alpha A_\beta \left(\frac{1}{4}\Phi^4
F^{\mu\nu}F_{\mu\nu} + \Phi\Phi^\mu_{;\mu}\right).
\end{eqnarray} where we have used the notation where a comma
indicates an ordinary derivative and a semicolon indicates covariant differentiation.

Using the fact that $\hat{R}_{\alpha\beta}=0$ gives now three sets
equations for the scalar $\phi$, the electromagnetic field $A_\mu$
and the $2D$ Ricci tensor $R_{\alpha\beta}$.  We again quote the
results of this transformation\cite{ref:Liu_Wesson}.
\begin{eqnarray*}
\Phi^\alpha_{;\alpha}&=&\frac{1}{4}\Phi^3F^{\alpha\beta}F_{\alpha\beta}\\
F^\lambda_{\alpha;\lambda}&=&-3\Phi^{-1}\Phi^\lambda F_{\lambda\alpha}\\
R_{\alpha\beta}&=&-\frac{1}{2}\Phi^2F_{\lambda\beta}F_\beta^\lambda+\Phi^{-1}\Phi_{\alpha;\beta}
\end{eqnarray*}

These equation can be interpreted as a wave equation for the
quantity $\Phi$, Maxwell's equations for the fields $A_\mu$ and the
Einstein field equations, the latter of which may be written

\begin{equation}
\label{eq:field equations} G^{\alpha\beta}\equiv
R^{\alpha\beta}-\frac{1}{2}g^{\alpha\beta}R=
8\pi\left(T_{em}^{\alpha\beta}+T_s^{\alpha\beta}\right)
\end{equation}
where the quantities $T_{em}^{\alpha\beta}$ and $T_s^{\alpha\beta}$
are effective energy momentum tensors given by
\begin{eqnarray*}
T_{em}^{\alpha\beta}&=&-\frac{1}{2}\Phi^2 \left(F^\alpha_\lambda
F^{\beta\lambda}-\frac{1}{4}g^{\alpha\beta}
F^{\mu\nu}F_{\mu\nu}\right)\\
T_s^{\alpha\beta}&=&\Phi^{-1}\left(\Phi^{\alpha;\beta}-g^{\alpha\beta}\Phi^\mu_{;\mu}\right)
\end{eqnarray*}
These results indicate that the definition of the electromagnetic
vector potential given in Equations~(\ref{eq:vector_potential}) is
consistent.
\section{Dirac's Equation}
 Given the definitions of the spin connection (\ref{eq:2D_spin_connection})
 and the electromagnetic field vector(\ref{eq:vector_potential})
 we now pick up the derivation of Dirac's equation from
 Equation~(\ref{eq:mode1}) which reads
\begin{eqnarray}
\label{eq:mode1_derivation1}
 \lefteqn{\sum_{\mu=1}^2 \tilde{\gamma}^\mu_0\left(\partial_\mu'
  - \imath\frac{m_1}{2}u_{3\mu,0} + \left[S^{-1}(\partial'_\mu
S)\right]_0\right)\tilde{\xi}_{1/2}
}\hspace{.5in}\\
 & & \mbox{}
+ \left(\frac{m_1}{2}+\imath\sum_{\nu=1}^2 u_{\nu3,0}
\partial_\nu' - \frac{m_1}{2}u_{33,0} -\imath
\left[S^{-1}(\partial'_3
S)\right]_0\right)\tilde{\xi}_{1/2}=0\nonumber
\end{eqnarray}
 Using
Equation~(\ref{eq:strain_decomposition}) the quantity $u_{3\mu}$ can
be written
\begin{eqnarray*}
u_{3\mu}&=&\epsilon_{3\mu}+\omega_{3\mu}+O(u^2)\\
 &=&\frac{1}{2}A_\mu+\omega_{3\mu}+O(u^2)
 \end{eqnarray*}
 and the quantity $u_{\mu3}$ is
\begin{eqnarray*}
u_{\mu3}&=&\epsilon_{3\mu}-\omega_{3\mu}+O(u^2)\\
 &=&\frac{1}{2}A_\mu-\omega_{3\mu}+O(u^2)
 \end{eqnarray*}
We also write the spinor field as
\[\Psi=\tilde{\xi}_{1/2}.\]
Equation~(\ref{eq:mode1_derivation1}) now becomes
\begin{eqnarray}
\label{eq:mode1_derivation2} \sum_{\mu=1}^2
\tilde{\gamma}^\mu\left(\partial_\mu'+
\imath\frac{m_1}{4}A_\mu - \Gamma_\mu\right)\Psi\\
&&\hspace{-2.3in}+ \left(\frac{m_1}{2}\left(1- u_{33,0}\right)
+\frac{\imath}{2}\sum_{\nu=1}^2
\omega_{3\nu}\partial'_\nu-\frac{\imath}{2}\sum_{\nu=1}^2
(A^\nu\partial'_\nu)+\left[S^{-1}(\partial'_3
S)\right]_0\right)\Psi=0\nonumber
\end{eqnarray}

\subsection{Mass Term}

The term $\frac{m_1}{2}(1-u_{33,0})$ can be written in terms of the
scalar $\Phi^2$ as

\[\frac{m_1}{2}(1-u_{33,0})=\frac{m_1}{4}\left(1-\Phi^2\right)+O(u^3)
\]
This scalar term can be interpreted as a mass term in
Equation~(\ref{eq:mode1_derivation2}) for a particle with mass $m$,
where $m$ is
\[
m=\frac{m_1}{4}\left(1-\Phi^2\right).
\]
Equation~(\ref{eq:mode1_derivation2}) now becomes
\begin{eqnarray}
\label{eq:mode1_derivation3} \sum_{\mu=1}^2
\tilde{\gamma}^\mu\left(\partial_\mu'+
\imath\frac{m_1}{4}A_\mu - \Gamma_\mu\right)\Psi+m\Psi\\
&&\hspace{-2.3in}+\left(\sum_{\nu=1}^2\frac{\imath}{2}
\omega_{3\nu}\partial'_\nu-\frac{\imath}{2}\sum_{\nu=1}^2
(A^\nu\partial'_\nu)+\left[S^{-1}(\partial'_3
S)\right]_0\right)\Psi=0\nonumber
\end{eqnarray}
With the exception of the last three terms, this equation has the
same form as Dirac's Equation~(\ref{eq:full_dirac}).  This is the
main result of this work which shows that a Dirac-like equation is
present in this system.

\subsection{Additional Interaction Terms}
The last three terms in Equation~(\ref{eq:mode1_derivation3}) can be
interpreted as interaction terms between the spinor field $\Psi$ and
the vector fields $A^\nu$, and $\omega_{3\nu}$, and the matrix field
$\left[S^{-1}(\partial'_3 S)\right]_0$. These terms do not normally
appear in Dirac's equation but the existence of these types of terms
is not forbidden in modern quantum field theories\cite{ref:Peskin}.

In addition to the equation given in (\ref{eq:mode1_derivation3}),
each of these fields satisfies additional constraint equations that
determines its dynamics.  The electromagnetic field satisfies
Maxwell's Equation and has been discussed in
section~\ref{sec:EM_equation}.  The interaction with the matrix
$\left[S^{-1}(\partial'_3 S)\right]_0$ results from the spin
connection in $3$ dimensions and its solution is determined from
Equation~(\ref{eq:aux_3D}).  An equation of motion for the field
$\omega_{3\mu}$ can be derived from
Equation~(\ref{eq:lagrangian_vector}).  If we take the divergence of
the vector $V^\rho$ from this equation we have
\[
\nabla^2u^\rho=-(2\lambda+2\mu)\frac{\partial\sigma}{\partial
x^\rho}+\frac{\partial \alpha}{\partial
x^\rho}+(\nabla\times\chi)^\rho
\]
which implies
\[
\nabla^2\frac{\partial u^\rho}{\partial
x^\nu}=-(2\lambda+2\mu)\frac{\partial^2\sigma}{\partial
x^\nu\partial x^\rho}+\frac{\partial^2 \alpha}{\partial
x^\nu\partial x^\rho}+\frac{\partial}{\partial
x^\nu}(\nabla\times\chi)^\rho
\]
We can now write the equation of motion of the field
$\omega_{\rho\nu}$ as
\[
\nabla^2\omega_{\rho\nu}\equiv\left(\frac{\partial u^\rho}{\partial
x^\nu}-\frac{\partial u^\nu}{\partial
x^\rho}\right)=\frac{\partial}{\partial
x^\nu}(\nabla\times\chi)^\rho-\frac{\partial}{\partial
x^\rho}(\nabla\times\chi)^\nu.
\]
We see that the scalar $\phi$ (and hence the spinor field $\xi$) is
completely determined by the scalars $\alpha$ and $\sigma$ in
Equation~(\ref{eq:lagrangian_vector}) and the field
$\omega_{\rho\nu}$ is completely determined by the vector field
$\chi$.

We will not attempt to identify these three additional terms with
known interactions, simply noting again that interactions of this
type are not forbidden to exist and that the central result of this
work is
 Equation~(\ref{eq:mode1_derivation3}) which shows that there is a
 Dirac-like equation that exists in a classical elastic solid.
\section{Higher Dimensions}
Throughout our derivation we assumed that we were working in three
dimensional space.  This was convenient since explicit solutions to
Laplace's equation could be obtained in terms of fractional
derivatives. The formalism however extends to any number of
dimensions\cite{ref:Cartan,ref:Brauer_Weyl}.  The main difference is
that there is no explicit solution of the spinors in terms of
fractional derivatives.  Nevertheless it should be noted that the
basic equations hold in higher dimensions and therefore the usual
four dimensional Dirac equations could in principle be derived by
starting with an elastic solid in $5$ dimensions.

\section{Conclusions}
We have taken a model of an elastic medium and derived an equation
of motion that has the same form as Dirac's equation in the presence
of electromagnetism and gravity.  We derived our equation by using
the formalism of Cartan to reduce the quadratic form of Laplace's
equation to the linear form of Dirac's equation. We further assumed
that one coordinate was compact and upon Fourier transforming this
coordinate we obtained a mass term and an electromagnetic
interaction term in the equations of motion. The formalism
demonstrates that the classical Einstein-Dirac-Maxwell equations can
be derived as the Equations of motion of the Fourier modes of an
elastic solid in the small strain approximation.

\end{document}